# On the Tidal Evolution and Tails Formation of Disc Galaxies

**M. Alavi** [(1)] and **H. Razmi** [(2)]

*Department of Physics, The University of Qom, 3716146611, Qom, I. R. Iran*

(1) m.alavi@stu.qom.ac.ir   (2) razmi@qom.ac.ir & razmiha@hotmail.com

## Abstract

In this paper, we want to study the tidal effect of an external perturber upon a disc galaxy based on the generalization of already used Keplerian potential. The generalization of the simple ideal Keplerian potential includes an orbital centripetal term and an overall finite range controlling correction. Considering the generalized form of the interaction potential, the velocity impulse expressions resulting from tidal forces are computed; then, using typical real values already known from modern observational data, the evolution of the disc including tidal tails formation is graphically investigated.

**Keywords**: galaxies; galactic interaction; tidal tail

# 1. Introduction

Our knowledge of tides and the tidal effects, as an old and familiar subject, corresponds to the tidal forces between the Earth and the Moon (Zeilik & Gregory 1998; Kay et al. 2013). The effect of tidal force upon the halo of comets in solar system, known as "Oort cloud" (Oort 1950) was investigated by Collins & Sari (2010). During recent years, the tidal forces have been proposed as the origin of many important natural phenomena at galactic scales, among which are the interaction of galaxies (Struck 1999), and clusters tidal stream (Renaud, Gieles & Boily 2011; Küpper, Macleod & Heggie 2008). The largest tidal stream has been already observed is in the local group known as Magellanic stream (Nidever et al. 2010) which is guessed to be due to the tidal interaction between the Magellanic Cloud and the Milky Way (Connors, Kawata & Gibson 2006). Holmberg (1940; 1941) and Zwicky (1959) had pointed out that the tails may be originated from the interaction of the galaxies. In 1972, using restricted three-body problem, and developing the study of Pfleiderer & Siedentopf (1961), Toomre & Toomre (1972) simulated the process of tail formation in the interaction of disk galaxies. In their model, considering the influence of the spin-orbit coupling, they investigated the tidal effects between two disk galaxies and concluded that the prograde encounters have stronger than retrograde ones confirming the old results by Holmberg (1941), Henon (1970), and Kozlov, Syunyaev & Éneev (1972). Keenan & Innanen (1975), using numerical method, and Read et al (2006), considering more improved analytical calculations, also confirmed that the prograde tides are dominated. The evidence for this fact that tidal evolution is strongest for the exactly prograde and weakest for the exactly retrograde orbit, using, has been recently reported by Lokas et al (2015) by means of *N*-body simulations. Tidal tails formation is one of the most outstanding consequences of the interaction of galaxies which is used in the study of a number of new subjects and discoveries such as the evolution of mergers (Bridge, Calberg & Sullivan 2010) and galaxies (Tal et al. 2009), distribution of dark matter (Bournaud, Duc & Masset 2003), and formation of tidal dwarf galaxies (Wetzstein, Naab & Burkert 2007). The Antennae (NGC4038/4039) (Hibbard et al. 2005) and the Superantennae (IRAS 19254-7245) (Charmandaris et al. 2002) are well-know systems with tidal tails. The remnants related to the tidal disruption of the satellites can form tails too (Choi, Weinberg & Katz 2007). Using impulsive approximation and considering the tidal effect as the perturbation in the orbit of stars in the disk, D'Onghia et al (2010) and Struck

& Smith (2012) have studied the system evolution and different properties of the tails with an analytical method. In this paper, considering a more realized effective interacting potential between the disc and the perturber than the simple ideal Keplerian potential already used, the tidal impulse is calculated and then the disk evolution snapshots for different values of angular momenta and the effective potential scale parameter are plotted.

## 2. Real Generalization of the Ideal Keplerian Potential

For the motion of a test particle in the gravitational field of a spherically symmetric mass $M$, with the Schwarzschild line element

$$ds^2 = \left(1 - \frac{2MG}{c^2 r}\right) c^2 dt^2 - \frac{1}{\left(1 - \frac{2GM}{c^2 r}\right)} dr^2 - r^2 d\Omega^2 \quad (1),$$

and the Lagrangian

$$\mathcal{L} = \left(1 - \frac{2GM}{c^2 r}\right) c^2 \dot{t}^2 - \frac{1}{\left(1 - \frac{2GM}{c^2 r}\right)} \dot{r}^2 - r^2 \left(\dot{\theta}^2 + \sin\theta^2 \, \dot{\varphi}^2\right) \quad (2),$$

using the corresponding Euler-Lagrange equations, it is found that the angular momentum per unit mass $h$ and the following $p$ parameters are constants of motion:

$$h \equiv r^2 \dot{\varphi} = constant, \quad p \equiv \left(1 - \frac{2GM}{rc^2}\right) \dot{t} = constant \quad (3).$$

where, without loss of generality, $\theta = \pi/2$ has been considered as the plane of the orbit.

Considering the total energy $E$

$$E \equiv \frac{1}{2}[P^2 - 1]c^2 = \frac{1}{2}\dot{r}^2 - \frac{GM}{r} + \frac{h^2}{2r^2} - \frac{GMh^2}{c^2 r^3} \quad (4),$$

The effective potential is found as (Rindler 2001; Hobson, Efstathiou & Lasenby 2006; Wang 2004)

$$\phi(r) = -\frac{GM}{r} + \frac{h^2}{2r^2} - \frac{GMh^2}{c^2 r^3} \quad (5).$$

In the above potential function, two generalizations consisting of a repulsive centripetal term and an attractive relativistic correction are seen relative to the simple ideal Keplerian potential $(-\frac{GM}{r})$.

Considering the real situation of the problem we are studying here and the corresponding observational data, one can simply check that the effect of the third term (the relativistic correction) is so small that it is better to be ignored; thus:

$$\phi(r) = -\frac{GM}{r} + \frac{h^2}{2r^2} \quad (6).$$

This effective potential is still an ideal potential because of its infinite range of influence. It takes a more real form by controlling its range with the following modification:

$$\phi(r) = \left(-\frac{GM}{r} + \frac{h^2}{2r^2}\right)e^{-\alpha r} \quad (7),$$

in which $\alpha$ is a constant of the order of $(\frac{1}{D})$ where $D$ is a distance up to 10-100 times of the disc extension.

The above form of the potential function is the same as in the real generalization of the ideal Coulomb potential in different problems of microphysics such as scattering problems in quantum physics, nuclear physics, and elementary particle physics known as the Yukawa potential.

## 3. The Method

In this section we consider a method as the same as what has been used by D'Onghia et al (2010) and Struck & Smith (2012) unless that we work with the modified potential (7).

For simplicity suppose the disk is located in *x-y* plane and the origin of the coordinate system is on its center. Assume the perturber passes through a straight line (Figure1) with the following position vector

$$\vec{r_p} = b\hat{i} + V_p t \hat{j} \quad (8).$$

Considering the disk galaxy as a collection of particles moving in circular orbits, the position vector for every star with respect to the origin of the coordinate system is

$$\vec{r_s} = r_s \cos(\Omega t + \varphi_0)\hat{i} + r_s \sin(\Omega t + \varphi_0)\hat{j} \quad (9).$$

The gravitational force acting on stars in the disk differs from the force at its center of mass and thus they experience tidal effects. Assuming the origin is located at the center of mass of the disk, the components of the tidal acceleration of each star are found as (Binney & Tremaine 2007)

$$a_{k(tidal)} = -\sum_j r_j \left(\frac{\partial^2 \phi}{\partial r_j \partial r_k}\right)_{\vec{r}=0} \quad (10),$$

in which $\phi$ is the effective generalized gravitational potential introduced in (7).

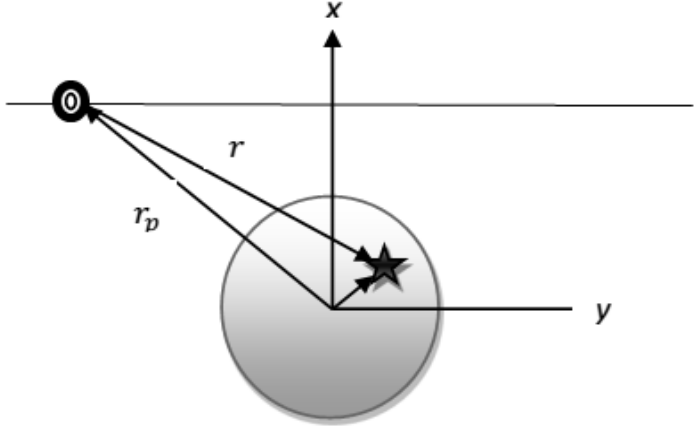

Figure 1: The geometrical configuration of the interaction between the disk galaxy and the perturber

Considering the velocity change due to the tidal force

$$\Delta v_k = -\int_{-\infty}^{\infty} \sum_j r_j \left(\frac{\partial^2 \phi}{\partial r_j \partial r_k}\right)_{\vec{r}=0} dt \quad (11),$$

its components in x-y plane are found as

$$\Delta v_x = \frac{r_s \cos\varphi_0}{b^2 V_p} \left\{ (2GM + \alpha h^2)[aK_1(a) + a^2(K_0(a) + K_1(a))] \right.$$
$$\left. - \left(\frac{\pi h^2}{2b}\right) e^{-a}[2 + 2a^2 + 3a] \right\} \quad (12),$$

$$\Delta v_y = \frac{r_s \sin\varphi_0}{b^2 V_p} \left\{ (-2GM - \alpha h^2)[a^2(K_0(a) + K_1(a))] \right.$$
$$\left. + \left(\frac{\pi h^2}{2b}\right) e^{-a}[2a^2 + a] \right\} \quad (13).$$

in them $b$ is the impact parameter, $a = \frac{\Omega b}{V_p}$ ($\Omega$ is the angular velocity of the stars), and $\varphi_0$ is the initial phase at $t=0$, and $K$ functions are the well-known modified Bessel functions. As we know, in the real world, the velocity change occurs in a finite time interval; thus, integrating from $t = -\infty$ to $t = \infty$ in the relation (11) is an ideal mathematical operation. Fortunately, using the generalized corrected potential introduced in (7), we can control this idealization and make our calculation be in accordance with the real world.

## 4. The Disk Evolution

The disk is assumed to consist of particles which are set in *150* rings. The radii of the rings are assumed to be about *0.056* to 1 times of the disc radius. The primary velocity of particles can be determined from the ideal Keplerian potential. In order to study the disk evolution due to the passage of the perturber, the velocity disruptions resulting from the tidal force should be added to the primary values in an algorithm in FORTRAN program. Typical real values for velocity, mass, and period are *V(R)* =*250 kms$^{-1}$*, *M(R)* =*1.5 × 10$^{11}$kg*, and *P(R)* =*250 Myr*. To compute the desired parameters and results, let work in the unit that mass *M=1*, Newton's gravitational constant *G=1*, the impact parameter *b=2.5*, and the perturber's velocity *V=7*. The outputs of the program corresponding to position coordinates of each particle have been plotted in two dimensions. Figure 2 shows the results of the evolution of the disk for two values of the effective potential constant parameter $\alpha=0.1$ and $\alpha=0.01$. Considering these two cases, the disc evolution occurs for the values of angular momentum $h$ in the interval $\sim 14$ to $20$. Clearly, for $\alpha$ smaller than or equal to $0.01$ (i.e. for the perturber's distance greater than or equal to $100$ times of the disk radius), there isn't any considerable evolution of the disk.

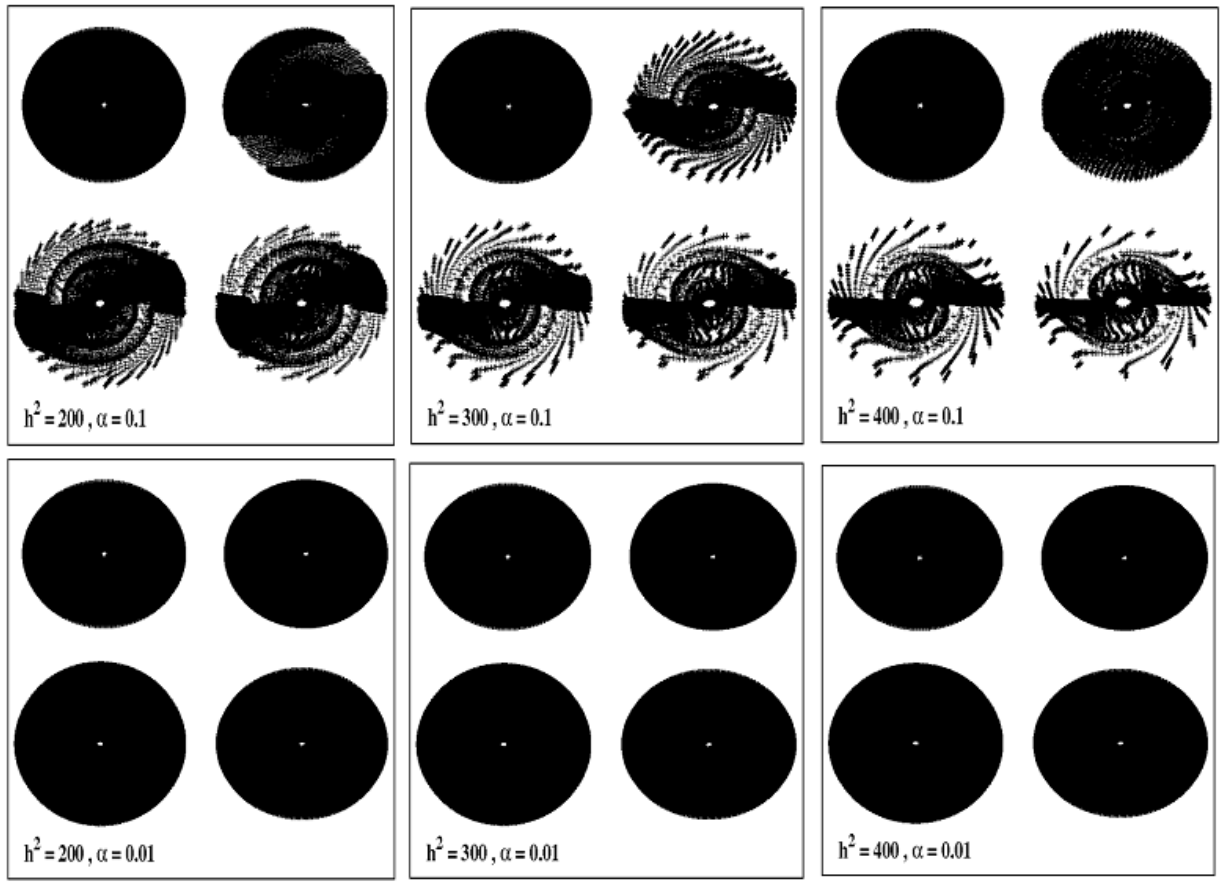

Figure 2: Evolution of the disk. Velocity = 7, impact parameter=2.5.

The limiting case $h=0$ and $\alpha=0$ (the ideal Keplerian potential model) has been illustrated in Figure 3.

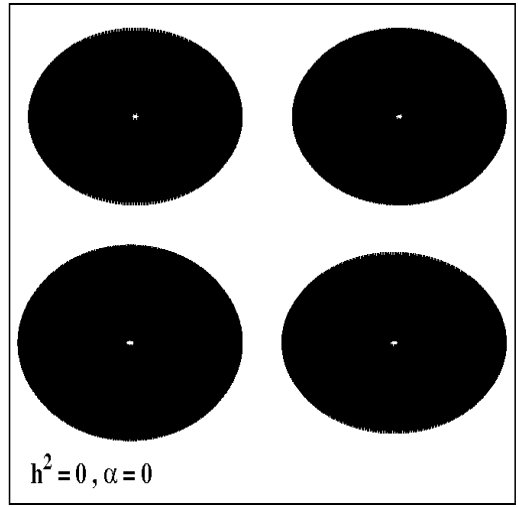

Figure 3: The disk evolution for the ideal Keplerian Potential.

## 5. Conclusion

After generalizing the simple ideal Keplerian potential by considering an orbital centripetal term and an overall finite range controlling correction, the large scale tidal effects on a disc galaxy have been studied. The resulting tidal evolution and deformation of the disc have been illustrated graphically in Figure 2 for typical real already known observational data. The resulting pictures and data can be used to explain how tidal tails are formed. Studying different cases of the resulting data and pictures for different values of the parameters and the constants appeared in the computations under consideration, one can fix the interval range of the influence of the angular momentum values and the effective potential controlling constant parameter $\alpha$. Comparing Figures 2 and 3 shows that there is a difference of degree of importance in considering the real generalized potential relative to the ideal Keplerian potential particularly in studying tails formation of the disk. The numerical values used in this paper are selected from real typical modern observational data so that lead us to nontrivial results.